# Nonreciprocal Bloch Oscillations in Magneto-Optic Waveguide Arrays

Miguel Levy and Pradeep Kumar

Department of Physics, Michigan Technological University, Houghton, Michigan 49931

ABSTRACT

We show that nonreciprocal optical Bloch-like oscillations can emerge in transversely magnetized waveguide arrays in the presence of an effective index step between the waveguides. Normal modes of the system are shown to acquire different wavenumbers in opposite propagation directions. Significant differences in phase coherence and decoherence between these normal modes are presented and discussed. Non-reciprocity is established by imposing unequal vertical refractive index gradients at the substrate/core, and core/cover interfaces in the presence of transverse magnetization.

OCIS codes: 130.2790, 230.7370, 000.1600

The present letter explores the properties of normal modes in nonreciprocal (NR) magneto-optic chirped waveguide arrays in the presence of a constant inter-waveguide wavenumber step. It is shown that the normal modes of the system display nonreciprocal propagation upon transverse magnetization to the propagation direction. The non-reciprocity of these states leads to NR Bloch-like oscillatory motion, albeit under conditions of approximate phase matching in opposite propagation directions. By embedding the system in a magneto-optic medium we demonstrate the possibility of optical modes akin to Wannier-Stark states, but having different properties upon propagation direction reversal. However, the presence of non-reciprocity in the system allows for NR normal-mode dephasing and the possibility of unidirectional Bloch oscillations.

Bloch oscillatory motion is a remarkable phenomenon first predicted by F. Bloch and C. Zener in the 1930's [1] consisting of oscillatory trajectories for particles subject to a constant force in periodic potentials. In optical systems this constant force is replicated by designing into the array a constant difference in waveguide mode index between adjacent waveguides [2-4]. A significant number of publications have discussed Bloch oscillations (BO) and the underlying Wannier Stark states in electronic and optical systems [2-7]. Work in this area has entailed theoretical and experimental demonstrations of optical BO and Wannier-Stark states. Possible applications to signal steering and the avoidance of nonlinearly-induced filamentation in high-power lasers have been suggested [2]. More recent work has examined optical BO in parity-time symmetric optical lattices and nonreciprocal phenomena as a result of gain, loss, or gain-loss modulation [5, 6].

Here we show that in the absence of gain or loss, normal modes of the waveguide array can exhibit different phase coherence lengths in opposite directions and even significantly different coherence and decoherence characteristics in the two directions. Our work extends previous treatments for two coupled-NR identical-waveguides to the case of coupled-multiple-waveguide arrays with different propagation constants in individual waveguides [8]. By imposing a uniform propagation-constant step between adjacent waveguides we make contact with BO and extend previous treatments to passive NR systems and different system-mode coherence effects in opposite directions.

We consider wave propagation in uniformly chirped waveguides in the presence of an in-plane transverse magnetic field. For each individual waveguide in the array, and assuming continuous wave propagation and no absorption, coupled-mode theory yields the following equation of motion for the modal amplitude $a_n^{f,b}$ of the $n^{\text{th}}$ waveguide

$$i\frac{da_n^{f,b}}{dz} + \delta\beta^{f,b} n a_n^{f,b} + \kappa^{f,b}(a_{n-1}^{f,b} + a_{n+1}^{f,b}) = 0. \qquad (1)$$

Here $f$ and $b$ denote the forward and backward directions. In Eq. (1) the wavenumber ($\beta = (2\pi/\lambda) \cdot n_{eff}$) of the $n=0$ guide has been separated out [2, 4]. $\lambda$ is the wavelength in vacuum, $n_{eff}$ is the waveguide-mode index, $\delta\beta^{f,b}$ is the difference in waveguide-mode wavenumber between adjacent waveguides, and $\kappa^{f,b}$ are the inter-waveguide coupling constants, in the forward and backward directions. The simultaneous constancy of parameters $\delta\beta^{f,b}$, hence the possibility of BO in opposite directions, is investigated below.

In normal mode formulation the light coupled into the central waveguide at the input facet can be described as a linear combination of normal modes, the $m^{\text{th}}$ mode propagation constant given by $\beta_0 + m\delta\beta$. Here $\beta_0$ is the propagation constant of the zeroth normal mode.

Refocusing (periodic Bloch-like oscillations) occurs when the normal modes recover an integer multiple of their initial phases, so that the Bloch oscillation period is $L_B = 2\pi/\delta\beta$, [2, 4].

In the waveguide array in Fig. 1 inset (a) the magnetization **M** is transverse to the propagation direction. The permittivity tensor is given by $\hat{\varepsilon} = \begin{pmatrix} \varepsilon & 0 & ig \\ 0 & \varepsilon & 0 \\ -ig & 0 & \varepsilon \end{pmatrix}$. Here $g$ is the gyrotropy parameter, propagation is along the z-direction and the magnetization points in the y-direction [8]. We take $\varepsilon$ and $g$ to be real numbers, making the model system lossless. This is a good approximation for magnetic garnets, such as bismuth/rare-earth-substituted iron garnets, in the near IR regime, where loss is small.

The dielectric tensor $\hat{\varepsilon}$ is spatially-dependent, having different values in the substrate, film (core) and cover regions of the waveguide. Specializing to modes that propagate in the z-direction (~exp(i$\beta$z)) and to quasi-transverse-magnetic (TM) modes ($H_y \gg H_x, H_z$, where **H** is the optical magnetic field component), one obtains [8],

$$\left(\varepsilon \partial_x \frac{1}{\varepsilon} \partial_x + \partial_y^2 - \beta^2 + \omega^2 \mu_0 \varepsilon_0 \varepsilon - \beta \frac{g}{\varepsilon} \partial_x + \beta \varepsilon \cdot \partial_x \frac{g}{\varepsilon^2}\right) H_y = 0 \qquad (2).$$

The linear terms in propagation parameter $\beta$ add up to $\beta \varepsilon H_y \partial_x (g/\varepsilon^2)$ yielding a net nonreciprocal propagation if the vertical gradients $\partial_x (g/\varepsilon^2)$ at the substrate/core and core/cover interfaces are different [8]. The change in propagation constant between forward and backward directions ($\Delta\beta^{(nr)} = \beta^f - \beta^b$) for system modes is given in perturbation theory by

$$\Delta\beta^{(nr)} = \frac{2\,\mathrm{Re} \int\int dxdy (\partial_x H_y) H_y^* (ig/\varepsilon^2)}{\int\int dxdy |H_y|^2 \varepsilon^{-1}}, \qquad (3)$$

where the superscript *nr* stands for non-reciprocal [8]. Modes are computed by three-dimensional semi-vectorial beam-propagation calculation with mode indices obtained using the correlation method [9].

The difference in inter-modal propagation constant between forward and backward directions is given by

$$\delta\beta^{f}_{m+1,m} - \delta\beta^{b}_{m+1,m} = \beta^{f}_{m+1} - \beta^{f}_{m} - \left(\beta^{b}_{m+1} - \beta^{b}_{m}\right) = \Delta\beta^{(nr)}_{m+1} - \Delta\beta^{(nr)}_{m} = \Delta(\Delta\beta^{(nr)})_{m+1,m}, \quad (4)$$

where $m$ labels the normal mode. In other words, the phase matching that produces Bloch oscillatory motion in a given direction is affected by the difference in the steps in NR propagation constant $\Delta(\Delta\beta^{(nr)})_{m+1,m}$. Assuming that Bloch-like oscillation conditions are satisfied in the forward direction, we obtain the following expression for backward propagation:

$$i\frac{da^{b}_{n}}{dz} + (\delta\beta^{f} - \Delta(\Delta\beta^{(nr)}_{n}))na^{b}_{n} + \kappa^{b}(a^{b}_{n-1} + a^{b}_{n+1}) = 0. \quad (5)$$

Here $\delta\beta^{f}$ and $\Delta(\Delta\beta^{(nr)}_{n})$ refer to differences between adjacent waveguides in the array. For an array of coupled waveguides the propagation constant difference $\delta\beta$ between individual waveguides coincides with those for the normal modes [4].

Figure 1 plots the calculated NR propagation constant difference $\Delta\beta^{(nr)}$ for each normal mode of the system using Eq. 3. Inset (b) displays the power distribution for the first normal mode. There is an approximate linear relation between these parameters, showing that $\Delta(\Delta\beta^{(nr)})_{m+1,m}$ is approximately constant, with a deviation from linearity of less than 2%. Under these conditions backward propagation yields Bloch-like oscillatory motion, with Bloch period given by $\frac{2\pi}{\delta\beta^f - \Delta(\Delta\beta^{(nr)})}$.

In designing the waveguide array we choose to vary the ridge waveguide thickness because the NR response in transversely-magnetized systems is most sensitive to thickness changes. In Fig. 1 the thickness has been adjusted (1.25 μm to 1.7 μm) to yield a constant $\delta\beta$ of about 6650 m$^{-1}$ between adjacent ridge waveguides. Ridge separations have also been selected to give a constant inter-waveguide coupling parameter $\kappa \sim 3650$ m$^{-1}$, with constant ridge width of 3 μm. The ensuing $\delta\beta$ produces a spatial Bloch period of ~ 945 μm and a lateral beam spread of less than 5 waveguides (~2.5 on each side) for light coupled into the center waveguide. The sequential change in ridge height introduces a nonreciprocal change in propagation constant $\Delta(\Delta\beta^{(nr)})$ of about 14 m$^{-1}$ between forward and backward directions between normal modes, for a typical value of $g \sim 0.004$ at $\lambda = 1.55$μm for bismuth-substituted iron garnets. Beam-propagation simulations (Fig. 2) show that a one-μm-wide beam launched into the center waveguide spreads out in both lateral directions but then returns to the center waveguide after a distance of about 945 μm. Upon modification of the inter-waveguide $\delta\beta$ by the nonreciprocal $\Delta(\Delta\beta^{(nr)}) \sim 14$ m$^{-1}$, a change in BO period of 2 μm is observed in the backward direction. The figure (Fig. 2) shows simulated propagation with different BO period in the backward direction, where the nonreciprocal effect has been artificially enlarged by a factor of 20 to make the effect visible in the scale of the figure. The correctness of the nonreciprocal effect discussed above was verified independently by magneto-optic full-wave finite-difference-time-domain simulation.

However, it is also possible to violate the conditions for Bloch oscillatory motion in one direction while maintaining a uniform wavenumber step in the opposite. In the previous example the inter-modal wavenumber difference $\delta\beta^f$ is much larger than the NR wavenumber shift $\Delta(\Delta\beta^{(nr)})$. A comparable magnitude in these two parameters is important to construct a system where the wavenumber step $\delta\beta^b$ is either cancelled out or changes value significantly across the waveguide array. This can be done by adjusting the waveguide widths and periodically reversing the sign of the gyrotropic parameter *g* between adjacent waveguides.

Sign reversal in *g* can be realized in Ga- or Al-substituted iron garnets by placing a silicon mask on the garnet film and subsequently annealing in nitrogen to create a compensation wall between the ridges [8]. Typical refractive indices for iron garnet films and a gadolinium gallium garnet substrate then yield $\Delta(\Delta\beta^{(nr)}) \sim \pm 1300$ m$^{-1}$ between adjacent waveguides. By introducing systematic changes in waveguide width, an equivalent and uniform inter-waveguide propagation difference $\delta\beta^f$ of ~ 1300 m$^{-1}$ in the forward direction can be produced. Under these conditions it is possible to construct an array with a constant inter-waveguide $\delta\beta^f$ and a strongly and spatially semi-periodic $\delta\beta^b$, oscillating by three orders of magnitude across the array, with significantly different propagation characteristics in the backward direction. Moreover, it is theoretically possible to attain a unidirectional breakdown in BO if the condition $\Delta(\Delta\beta^{(nr)}) = \delta\beta^f$ is satisfied throughout the array.

Applications of the nonreciprocal BO phenomena discussed above are expected to be rather general. These could include the juncture of normal optical effects with nonreciprocity, such as magneto-optically-controlled bi-directional signal steering and switching, all-optical nonreciprocal switching, rerouting and channel reconfiguring. Extension of the NR BO effect to semiconductor materials is also possible. Calculations performed by the authors also show that silicon-based waveguides with sputter-deposited magneto-optic cover exhibit significant NR BO.

In summary, we demonstrate the existence of nonreciprocal Bloch oscillations in chirped gyrotropic waveguide arrays with transverse magnetization. A nonzero phase coherence length difference between counter-propagating normal modes is obtained. It is shown that an array can be constructed with a constant wavenumber step in one propagation direction that simultaneously violates the conditions for standard Bloch oscillatory motion in the opposite.

This material is based upon work supported by the National Science Foundation under Grant No. 0856650. The authors thank MJ Steel for the FDTD simulation.

FIGURE CAPTIONS

FIG. 1. (Color online). Plot of nonreciprocal $\Delta\beta^{(nr)}$ versus normal mode $\beta$ showing a nearly uniform $\Delta(\Delta\beta^{(nr)})$ between consecutive normal modes. The insets show (a) a schematic depiction of the waveguide array highlighting the effective index progression between adjacent waveguides, and (b) the power distribution of the first normal mode of the array.

FIG. 2. (Color online) Beam-propagation simulations of (a) Bloch mode oscillations for the waveguide array with film index 2.35 and substrate index 2.25, showing a BO period of ~ 945 µm for a 1µm-wide input beam. (b) Simulated Bloch oscillations in forward, and (c) backward directions. The strength of $\Delta(\Delta\beta^{(nr)})$ has been artificially enhanced by a factor of twenty to highlight the difference in BO period shown with dashed lines.

FIGURES

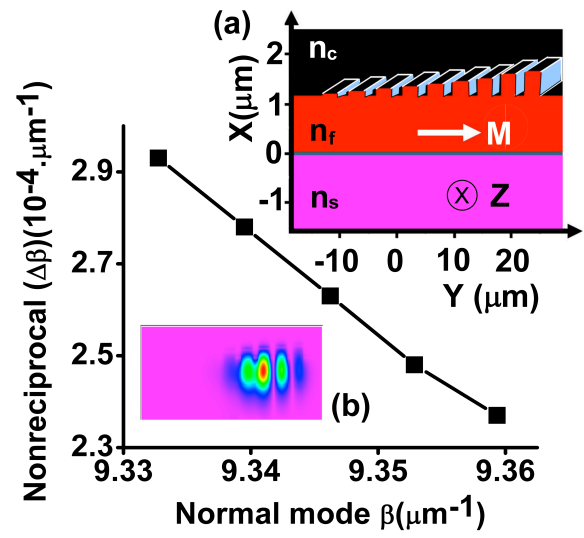

Fig. 1

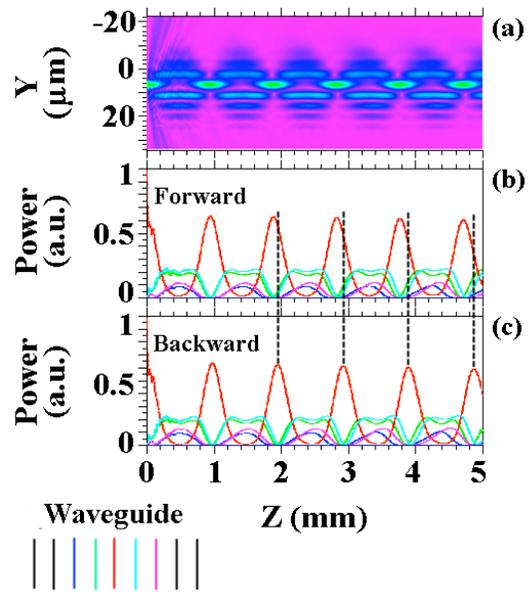

Fig. 2